\documentclass[prd, 10pt,aps,nofootinbib, floatfix, notitlepage,twocolumn,longbibliography]{revtex4-1}

\usepackage{amsmath,amsfonts,amssymb}
\usepackage{mathrsfs}
\usepackage{graphicx}
\usepackage[english]{babel} 
\usepackage{url} 
\usepackage{color}
\usepackage{bbold}
\usepackage{adjustbox}
\usepackage[utf8]{inputenc} 
\usepackage[colorlinks=true,citecolor=blue,urlcolor=blue]{hyperref}

\newcommand{\be}{\begin{equation}}
\newcommand{\ee}{\end{equation}}
\newcommand{\bes}{\begin{equation*}}
\newcommand{\ees}{\end{equation*}}

\newcommand{\LCDM}{$\Lambda$CDM}

\newcommand{\DA}{\text{DA }}
\newcommand{\PSKM}{\hyperlink{cite.Poulin:2018cxd}{P18}}

\begin{document}

\title{Thermal Friction as a Solution to the Hubble Tension}
\author{Kim V. Berghaus$^1$}
\email{kbergha1@jhu.edu}
\author{Tanvi Karwal$^{1,2}$}

\affiliation{$^1$Department of Physics and Astronomy, Johns
                              Hopkins University, 3400 N.\ Charles St., Baltimore, MD 21218, United States}
\affiliation{$^2$Center for Particle Cosmology, 
                Department of  Physics and Astronomy,
                University of Pennsylvania, 
                209 S. 33rd St., Philadelphia, PA 19104, 
                United States}

\begin{abstract}
A new component added to the standard model of cosmology that behaves like a cosmological constant at early times  and then dilutes away as radiation or faster can resolve the Hubble tension. We show that a rolling axion coupled to a non-Abelian gauge group 
 exhibits the behavior of such an extra component at the background level and can present a natural particle-physics model solution to the Hubble tension. We compare the contribution of this bottom-up model to the phenomenological fluid approximation and determine that CMB observables sensitive only to the 
background evolution of the Universe are expected to be similar 
in both cases, strengthening the case
for this model to 
provide a viable solution to the Hubble tension.
\end{abstract}

\date{\today}

\maketitle

\section{Introduction}
\label{sec:intro}

The tremendously successful standard model of cosmology assumes a flat universe, cold dark matter (CDM) and cosmological-constant dark energy $\Lambda$. 
This \LCDM model correctly describes numerous observables including the the complex structure of the cosmic microwave background (CMB) spectra \cite{Ade:2015xua,Aghanim:2018eyx}.
However, its predictions for the current rate $H_0$
of expansion of the Universe based on the CMB
are discrepant with the most precise direct measurements 
in the local universe at $>4\sigma$
\cite{Riess:2018byc,Freedman:2017yms,Verde:2019ivm,Wong:2019kwg}. 
With no obvious systematic cause in sight \cite{Efstathiou:2013via,Addison:2015wyg,Aylor:2018drw,Macpherson:2018akp,Camarena:2018nbr,Jones:2018vbn,Kenworthy:2019qwq,Follin:2017ljs,Spergel:2013rxa,Feeney:2017sgx,Pandey:2019yic}, 
this worsening tension has inspired many theorists to 
postulate new physics beyond the \LCDM\ model \citep[for e.g., and references therein]{Freedman:2017yms,10.1093/mnras/sty418, Blinov:2019gcj, Knox:2019rjx,Desmond:2019ygn}. 
However, few solutions exist \cite{Poulin:2018cxd,Agrawal:2019lmo,Lin:2019qug,Smith:2019ihp,Kreisch:2019yzn, Blinov:2019gcj} 
that simultaneously resolve the 
Hubble tension while also providing a good fit to all observables. 

One of the more successful solutions
is the addition of an early dark energy (EDE) component 
\cite{Poulin:2018cxd,Karwal:2016vyq, Agrawal:2019lmo, Lin:2019qug}, 
disjoint from the late-time dark energy. 
This component behaves like a cosmological constant at early times, 
then dilutes away as fast or faster than radiation at 
some critical redshift $z_c$, localizing its influence on cosmology 
around $z_c$. 
It increases the pre-recombination expansion rate, decreasing 
the size $r_s$ of the sound horizon. 
The CMB inference of $H_0$ is based on $r_s$ and its angular 
size $\theta_*$ on the surface of last scatter. 
Precise observations of $\theta_*$ combined with 
a \LCDM-based deduction of $r_s$ determine
$H_0$ as $\theta_* \sim r_s H_0$. 
Hence, a theory that predicts a smaller $r_s$ 
also infers a greater $H_0$ to preserve 
the precisely measured $\theta_*$, alleviating the Hubble tension. 
It was proposed as a phenomenological solution, the dynamics of
which could emerge from various particle-physics models
\cite{Agrawal:2019lmo,Lin:2019qug,Smith:2019ihp,Alexander:2019rsc,Niedermann:2019olb,Sakstein:2019fmf}.

In this paper, we present a dynamical particle-physics model 
that could solve the Hubble tension, 
which at the background level, mimics the evolution of early dark energy. 
This model, the "dissipative axion" (DA), is presented in Sec. \ref{sec:model}.
Although we leave the details of the perturbations of this model 
to future work, in Sec.~\ref{sec:background}, we argue why 
the background dynamics of this model are promising
and indicate that the \DA can form an extra dark energy component that resolves the Hubble tension. 
We conclude in Sec.~\ref{sec:concl}, where we
discuss the broader implications of this model and 
the way forward.

\section{Model}
\label{sec:model}

We add a pure dark non-Abelian gauge group
[$\text{SU}(\text{2})$] and an axion $\phi$ 
to the Standard Model particle content. 
The dark gauge bosons interact with $\phi$ via a CP-odd coupling, 
\begin{equation} \label{int}
    \mathcal{L}_{\text{int}} = 
    \frac{\alpha}{16 \pi}
    \frac{\phi}{f}
    \tilde{F}^{\mu\nu}_a F^a_{\mu\nu} \,,
\end{equation}
where $F^a_{\mu\nu}$ ($\tilde{F}^a_{\mu\nu} 
= \epsilon^{\mu\nu\alpha\beta} F^a_{\alpha \beta}$) 
is the field strength of the dark gauge bosons and 
$\alpha =\frac{g^2}{4 \pi}$, where $g$ is the gauge coupling 
of the dark group. 
The dark sector is decoupled 
from the standard model. We give the axion, which is displaced from its minimum, a simple UV-potential \footnote{The IR potential from the confining group is rapidly suppressed at temperatures above the confining scale and we have checked that its contribution is sub-dominant for our parameters.},
\begin{eqnarray}
    V(\phi) = \frac{1}{2}m^2 \phi^2 \,.
    \label{eq:potential}
\end{eqnarray}
\enlargethispage{\baselineskip}
This potential intuitively illustrates the dynamics of our model,
as the axion is essentially an overdamped harmonic oscillator. 
The interaction term $\mathcal{L}_{\rm int}$ 
adds an additional friction 
$\Upsilon(T_{\text{dr}})$ 
to the equation of motion,
dissipating energy through the production 
of dark radiation $\rho_{\text{dr}}$ which is 
comprised of dark gauge bosons, where $T_{\text{dr}}$ is the temperature of the dark radiation. 
In the small coupling limit ($\alpha \ll 1$), 
$m \ll \alpha^2 T_{\text{dr}}$, and this friction 
can be inferred from the sphaleron rate for a pure non-Abelian gauge group
\cite{PhysRevD.43.2027, Moore:2010jd, Laine:2016hma} 
and scales as
\begin{equation}
    \Upsilon(T_{\text{dr}}) = \kappa \alpha^5  \frac{T_{\text{dr}}^3}{f^2}    \,,
\end{equation}
where $\kappa$ is an O(10) number\footnote{For a general SU(N) $\kappa$ increases with $N$. For details see \cite{Moore:2010jd}.} with 
weak dependence on $\alpha$ and $f > T_{\text{dr}}$. 
The following equations of motion then describe 
the homogeneous evolution of the axion-radiation system: 
\begin{eqnarray} \label{eom}
    \ddot{\phi} +\left(3H + \Upsilon \left(T_{\text{dr}} \right) \right) \dot{\phi} +m^2 \phi &=& 0 
    \nonumber \\ 
    \dot{\rho}_{\text{dr}} +4H \rho_{\text{dr}}  &=& \Upsilon(T_\text{dr}) \dot{\phi}^2   
    \label{eq:eqs_of_motion}
\end{eqnarray}
where $\rho_{\text{dr}} = \frac{\pi^2}{30} g_{*} T^4_{\text{dr}}$
and $g_* = 7$ denotes the relativistic degrees of freedom
in the new dark sector. ($g_* = 2(N^2-1)+1$ for a general SU(N), where the factor of $2$ accounts for two gauge boson polarizations per gauge boson ($N^2-1$) and the axion contributes $1$ additional degree of freedom.)

In the original EDE work, an oscillating scalar field 
subject only to Hubble friction had been proposed, 
whose energy must dilute like radiation or faster 
after the field becomes dynamical 
in order to diminish the Hubble tension.  
This requirement places rigid demands on the 
scalar-field potential 
$V \propto \left( 1 - \cos \frac{\phi}{f} \right)^n$ 
considered by \cite{Smith:2019ihp} (or $V \propto \phi^{2n}$ as in \cite{Agrawal:2019lmo})
with $n \geq 2$. 
These potentials do not easily emerge from a 
UV-complete theory without extreme fine-tuning.
Other proposed phenomenological EDE candidates \cite{Lin:2019qug} 
have similar fine-tuning issues.  

In our \DA model, the particle-production friction
$\Upsilon \gg m, 3H$, overdamps the motion of the scalar field.
Thus, because the field is not oscillating, 
its dynamics are not sensitive to the potential $V(\phi)$. 
Instead, the friction $\Upsilon$ extracts energy from 
the scalar field into the dark radiation, 
which automatically dilutes away as $a^{-4}$. 

We approximate the solution to the equation of motion 
Eq.~\eqref{eom} as 
\begin{equation} \label{phi}
    \phi(z) \approx \phi_0 e^{-\frac{ 
            m^2}{H(z) \Upsilon(z)}  } \,,
\end{equation}
which is the solution to  
an overdamped oscillator where we approximated 
$t \simeq {H(z)}^{-1}$. 
Equation \eqref{phi} illustrates that the \DA
begins to roll faster when $\frac{\Upsilon(z_d)}{m^2} \equiv H(z_d)$, 
where $z_d$ denotes the redshift at which the axion field 
becomes dynamical. 
At high redshifts ($z \gg z_d$) the axion is slowly rolling, building up to a steady-state temperature on time scales of order $\Upsilon^{-1}$ in the dark sector,
\begin{equation} \label{ssT}
    T_{\text{dr}}(z) \approx \left(\frac{m^4 f^2 \phi^2(z)}{2 \frac{\pi^2} {30} g_{*} \kappa \alpha^5 H(z)}  \right)^{\frac{1}{7}} \,,
\end{equation}
by continuously extracting 
energy from the rolling field \cite{Berghaus:2019whh}. As the field begins to roll faster, the temperature $T_{\text{dr}}$ 
in the dark sector rises steadily and the field continuously
dumps its energy into the dark radiation bath. 
However, due to the weak dependence of the temperature 
on the background quantities, this change is O(1). 
Therefore, approximating the friction $\Upsilon(z)$ 
as roughly constant does not change the qualitative behavior 
of our model at the background level, 
as we discuss in more detail in Sec.~\ref{sec:background}. 
Eventually, as the axion energy depletes, the source term 
$\Upsilon \dot{\phi}^2$ becomes smaller than 
$4H \rho_{\text{dr}}$, leading to a decrease 
in temperature $T_{\rm dr}$ until $\Upsilon \dot{\phi}^2$
becomes negligible and the dark radiation 
dilutes away as $a^{-4}$. 

The generation of a steady-state temperature is independent of the presence of an initial dark temperature, as even starting with temperature fluctuations of the order of Hubble is sufficient to rapidly build up to the temperature in Eq. \eqref{ssT} \cite{Berghaus:2019whh}. 
Indeed, the main features of the DA
are universal in the presence of 
any large friction [$\Upsilon \gg H(z)$]
for $\Upsilon \propto T^p$ 
with $p < 4$. 
The minimal model presented here has been explored in more detail \cite{Berghaus:2019whh} in the context of warm inflation \cite{Berera:1995ie, Berera:1995wh, Berera:1999ws, Berera:1998px, Berera:2008ar, Bastero-Gil:2016qru}.

\section{Background Dynamics}
\label{sec:background}

\begin{figure}
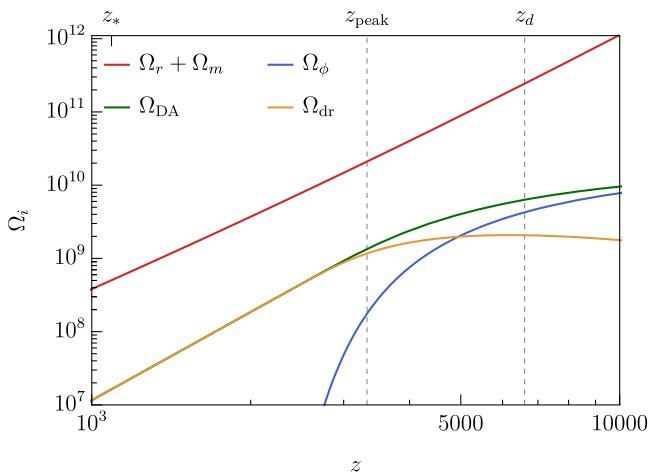

    \adjustimage{width=0.48\textwidth,center}{"Fig1".pdf}
    \caption{
    The fractional energy densities 
    $\Omega_i = \rho_i / \rho_{\rm crit}$ of 
    the different components in the DA
    and those in a \LCDM\ universe, where $\rho_{\rm crit}$ is the critical density today.   
    The total \DA contribution (green) is a sum of 
    its sub-components. 
    At early times ($z \gg z_d$), the energy density 
    $\Omega_{\phi}$ in the scalar field (blue) is roughly constant 
    and the dark radiation component $\Omega_{\text{dr}}$ (yellow) is subdominant.
    At intermediate times $(z_{\text{peak}} < z < z_d)$, the dark radiation
    $\Omega_{\text{dr}}$ transitions to become dominant as $\Omega_{\phi}$ drops.
    Shortly after $T_{\text{dr}}$ reaches a maximum, 
    the total fractional \DA energy density
    peaks at redshift $z_{\text{peak}}$.
     } 
     \label{fig:tot_energy_density}
\end{figure}

Having laid the groundwork for the background evolution 
of the \DA, 
we turn to its ability to mimic EDE
and draw comparisons with the best-fit parameters 
of Ref.~\citep[henceforth labeled \PSKM]{Poulin:2018cxd}. 
The particle setup in Sec.~\ref{sec:model}
results in a rolling scalar field that behaves 
like a cosmological constant at early times plus 
a dark radiation component. 
The total contribution $\rho_{\text{DA}}$ 
to an EDE-like component 
is then given by their sum
\begin{eqnarray} \label{EDE}
    \rho_{\text{DA}} (z) 
    = \rho_{\phi}(z)  + \rho_{\text{dr}}(z) ,
\end{eqnarray}
where $\rho_{\phi}(z) \approx \frac{1}{2}m^2 \phi^2(z)$\footnote{The 
kinetic energy component of $\phi$ is negligible 
due to the large friction term.}. 
At very early times, the radiation component is sub-dominant 
and $\phi$ is essentially frozen, acting like a 
cosmological constant giving
\begin{equation}
    \rho_{\text{DA}} (z \gg z_d) \approx \frac{1}{2}m^2 \phi^2_0 \,, 
\end{equation}
which is a function of only the axion potential 
and its initial conditions. 
Sometime after the axion thaws ($z < z_d$),  
the dark radiation becomes the dominant contributor to EDE
as illustrated in Fig.~\ref{fig:tot_energy_density}. 
The DA constitutes a total fraction, 
\begin{eqnarray}
    f_{\text{DA}}(z) = \frac{\rho_{\text{DA}}(z)}
             {\rho_m(z) + \rho_r(z) + \rho_{\text{DA}}(z)} 
\end{eqnarray}
of the energy density of the Universe, 
where $\rho_m$ and $\rho_r$ denote the matter 
and radiation densities. 
This fraction reaches a maximum at $z_{\text{peak}}$. 
Relating this to the "critical redshift" $z_c$ of the EDE 
as defined in \PSKM, 
their best fit $z_c=5345$\footnote{The posteriors for EDE parameters
in \PSKM\ are non-Gaussian. The best-fit parameters quoted here 
therefore do not correspond to their mean values, and we hence 
do not include errors on these quotes. }
for the EDE that dilutes as radiation, 
which corresponds to $z_{\text{peak}} =3322$.
Roughly at this time, the source term $\Upsilon \dot{\phi}^2$
in Eq.~\eqref{eom} becomes negligible, and 
the dark radiation dilutes away as $a^{-4}$ as shown 
in Fig.~\ref{fig:tot_energy_density}. 

By approximating the friction $\Upsilon(z_{\rm peak}) = \Upsilon_0$ 
as a constant, we illustrate how to estimate $z_{\rm peak}$ analytically. In this limit, the approximation for the temperature of the dark radiation simplifies to  
\begin{equation} \label{Trd}
    T_{\text{dr}}(z > z_{\text{peak}}) \simeq \left(\frac{m^2 \phi(z)}{2 \sqrt{\frac{\pi^2}{30} g_{*} H(z)\Upsilon_0} } \right)^{\frac{1}{2}} ,  
\end{equation}
which, using Eqs.~\eqref{phi} and \eqref{EDE}, 
allows us to approximate
$f_{\text{DA}}$ as an analytical function in $z$, 
\begin{equation} \label{fDA}
    f_{\text{DA}}(z \geq z_{\text{peak}}) \simeq
        \frac{e^{-\frac{2m^2}{H(z)\Upsilon_0}}
            \frac{1}{2} m^2 \phi^2_0 
            \left(1 + \frac{m^2}{2H(z)\Upsilon_0} \right)}
            {\rho_m(z) +\rho_r(z)}
            .
\end{equation}
Solving $\frac{df_{\text{DA}}}{dz}|_{z_\text{peak}}=0$, 
and assuming that the peak lies close to 
matter-radiation equality, we can approximate $z_{\rm peak}$ as 
\begin{equation} \label{zpeak}
    z_{\text{peak}} \simeq \left(\frac{1}{2\sqrt{\Omega_m}} \frac{ m^2}{H_0 \Upsilon_0} \right)^{\frac{2}{3}} ,
\end{equation}
where $\Omega_m$ is the fractional matter density today and 
$z_{\rm peak}$ is now
dependent only on $\frac{\Upsilon_0}{m^2}$. 
Equations \eqref{Trd}$-$\eqref{zpeak}  
demonstrate how the physical observables depend exclusively 
on $\frac{\Upsilon}{m^2}$, which sets the time scale 
at which the axion becomes dynamical, 
and $\frac{1}{2}m^2 \phi_0^2$ which scales 
the total amount of early dark energy. 
Therefore, at the background level, we effectively introduce 
only two new parameters beyond \LCDM, 
but expect the perturbations to depend 
on more than just these two parameters.
Including the full temperature dependence of the friction at the background level
requires solving the coupled differential Eq. \eqref{phi}
numerically by specifying an initial condition
$\frac{\Upsilon(z_i)}{m^2}$ at some $z_i$, 
increasing the effective number of background parameters to three. 
While this does not have a significant impact on 
the qualitative behavior of the \DA system, 
it does change $\frac{\Upsilon(z_i)}{m^2}$,
and $\frac{1}{2}m^2 \phi_0^2$ by O(1) when keeping
$z_{\text{peak}}$ and $f_{\text{DA}}(z_{\text{peak}})$ fixed.

\begin{figure}
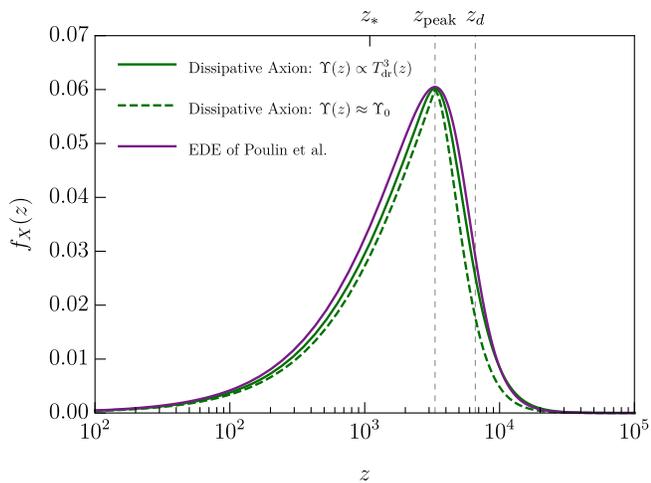
 
    \adjustimage{width=0.48\textwidth,left}{"Fig2".pdf}
    \caption{We compare the fractional early dark energy density
    of the full temperature dependent \DA model
    [$\Upsilon(z) \propto T^3_{\text{dr}}$, solid green] 
    with the semi-analytical approximations in equation
    \eqref{fDA} and \eqref{rhoda}, treating the friction 
    as constant  [$\Upsilon(z) \approx \Upsilon_0$ dashed green]
    and the EDE fluid approximation of an oscillating scalar field 
    from Poulin et. al. \cite{Poulin:2018cxd} (purple). 
    This plot uses the $n=2$ EDE best-fit parameters  
    [$z_{\text{c}} = 5345, 
    f_{\text{EDE}}(z_{\text{c}}) =0.044$ 
    which corresponds to $z_{\text{peak}} = 3322,\,
    f_{\text{EDE}}(z_{\text{peak}}) =0.060$] 
    and dissipative axion parameters
    $\frac{\Upsilon(z_{\text{peak}})}{m^2} = 1.3 *10^{36}\, \text{GeV}^{-1}$ 
    $\Big(\frac{\Upsilon_0}{m^2} = 5.7 *10^{36}\,\  
    \text{GeV}^{-1}\Big)$, and 
    $\frac{1}{2}m^2 \phi^2_0 = 0.55 \,\ \text{eV}^4$ 
    $\Big(\frac{1}{2}m^2 \phi^2_0 = 0.21 \,\  \text{eV}^4\Big)$ 
    for the temperature dependent (independent) \DA model. 
    }
    \label{fig:frac_energy_density}
\end{figure}

For redshifts smaller than $z_{\text{peak}}$, 
the early dark energy is dominated by the 
radiation component which dilutes as: 
\begin{eqnarray} \label{rhoda}
    \rho_{\text{DA}} (z < z_{\rm peak}) \simeq 
        \rho_{\text{dr}}(z_{\rm peak}) 
        \left(\frac{1+z}{1+z_{\rm peak}}\right)^4 .
\end{eqnarray}
The fractional energy density $f_{\rm DA}$ is then peaked at $z_{\rm peak}$,
as shown in Fig.~\ref{fig:frac_energy_density}. 
Our proposed model hence mimics the EDE proposed in 
\PSKM\ with $n = 2$, which resolves the Hubble tension.

The primary difference between the two models 
at the background level is a narrower peak 
for the \DA (the effect being more pronounced 
for the constant friction approximation), 
as seen in Fig.~\ref{fig:frac_energy_density}. 
Based on this, we explore the expected differences 
between the background observables of the two models. 
In particular, we discuss the impact on CMB observables 
that capture the important features of the full CMB spectrum, 
but depend only on the background evolution 
of the Universe \cite{Hu:2000ti,Knox:2019rjx,Poulin:2018cxd}. 
These are the size $r_s$ of the sound 
horizon, the ratio $r_{\rm damp}/r_s$ 
of the damping scale to the sound horizon, 
the height of the first peak and the horizon size
at matter-radiation equality. 

As our model adds more radiation to the Universe, 
we naively expect the redshift of matter-radiation 
equality to shift. Quantifying this shift correctly 
requires a full Markov chain Monte Carlo (MCMC) to allow other cosmological 
parameters, in particular the physical density 
$\omega_{\rm cdm}$ of cold dark matter to compensate 
for some or all of the shift. 
We expect that the results of the MCMC will pull our 
posteriors in a direction that minimizes change
to $z_{\rm eq}$.
We hence leave further discussion of changes 
to $z_{\rm eq}$ for future work. 
We expect an increase in $\omega_{\rm cdm}$ to 
similarly compensate for a change to the height 
of the first CMB peak. 
Such an increase was observed by \PSKM\ for EDE - 
the best-fit $\omega_{\rm cdm}$ increases by $\sim 9\%$ 
in the $n=2$ EDE cosmology relative to \LCDM. 
To compare, their maximum $f_{\rm EDE} \leq 7\%$.  
Moreover, the dark radiation peaks during matter-domination, further
minimizing the effect of adding dark radiation to the Universe. 
Consequently, in this paper, we limit 
our comparisons of the two models to investigating 
the effects of the sharper peak in $f_{\rm DA}$. 

We first note that a slight narrowing of the peak
of $f_{\text{DA}}$ relative to $f_{\rm EDE}$ has 
minimal impact on the recombination redshift $z_*$. 
This was verified using a modified version of the 
equation of state parametrization of the EDE 
of \PSKM, similar to Ref.~\cite{Lin:2019qug}, 
sharpening the peak in $f_{\rm EDE}$ and calculating $z_*$ 
with the CLASS cosmology code \cite{lesgourgues2011cosmic, Blas_2011}. 
As $z_*$ is a background quantity, and $f_{\text{DA}}$
is nearly identical to a narrower $f_{\rm EDE}$, we expect 
$z_*$ for the \DA to be similar 
to the EDE scenario. 
Then, the main change to $r_s$ comes not from 
the limits of its integral, but the integrand, 
specifically, the expansion rate. 
Knowing how the expansion rate for the \DA differs from EDE, we can calculate $r_s$
by fixing the background cosmology to the
best fit of the $n=2$ EDE of \PSKM, 
and the DA parameters such that 
the temperature dependent (independent)
$z_{\rm peak}$ and $f_{\DA}(z_{\rm peak})$ match the best-fit EDE 
(values specified in the caption of
Fig.~\ref{fig:frac_energy_density}), giving 
\begin{eqnarray}
    r_s(z_*) 
        = \int^{\infty}_{z_*} dz \frac{c_s(z)}{H(z)} 
        = 140.0 \, (140.1)\, \text{Mpc} 
        \,,
\end{eqnarray}
compared to $r_s = 139.8 \,$Mpc in \PSKM.
Here, $c_s(z)$ is the speed of sound in plasma and 
the \DA enters into the expansion rate $H(z)$. 
This is well within $1\sigma$ of the $r_s$ 
in the best-fit EDE scenario of \PSKM\ for $n=2$, for which the 
best-fit Hubble constant increases 
to $H_0 = 71.1$ km/s/Mpc. 
This along with a larger error on $H_0$ resolves the tension in the EDE case. 
As the CMB inferences of $r_s$ and $H_0$ are degenerate, 
with a reduced $r_s$ that matches \PSKM\ in the DA model, we similarly expect a high $H_0$ that will significantly ease the Hubble tension, if not resolve it.

For $r_{\rm damp}$, we expect a smaller change still, as
the integral for $r_{\rm damp}$ 
is sharply peaked close to recombination and 
less sensitive to the expansion rate $\sim z_{\rm eq}$. 
While the change in $r_s$ is absorbed by $H_0$, 
thereby diminishing the Hubble tension, 
changes to $r_{\rm damp}/r_s$ can be absorbed by
the tilt $n_s$ of the primordial power spectrum as noted
by Refs.~\cite{Poulin:2018cxd,Knox:2019rjx}. 

Another requirement of EDE models that succeed in resolving 
this discrepancy is an effective sound speed $c_s^2 < 1$
of perturbations in the new component
\cite{Lin:2019qug,Smith:2019ihp,Agrawal:2019lmo}. 
This in part led to the success of
Refs.~\cite{Smith:2019ihp,Poulin:2018cxd}.  
The DA model consists of a scalar field ($c_s^2 = 1$) and 
dark radiation ($c_s^2 = 1/3$) \cite{Hu:1998kj}. 
Although the coupling between the two components complicates matters, 
as $\rho_{\phi} < 20\%$ at $z_{\rm peak}$, the rest
of the energy density being made up of dark radiation, 
naively, we expect $c_s^2$ for the DA to be 
between $1/3 < c_s^2 < 1$. Here, we simply seek to motivate the relevance of this model
as a particle theory solution to the Hubble tension, 
and leave the exploration of perturbations to subsequent work. 
As the \DA model
produces a value for $r_s$ extremely close to
the EDE value, and little to no difference 
is expected in $r_{\rm damp}$ between the two models, 
these expectations coupled with the predicted increase
in $\omega_{\rm cdm}$ make the \DA a 
promising theoretical model to deliver the extra early dark energy component
that can resolve the Hubble tension.

\section{Discussion}
\label{sec:concl}

In this paper, we propose the \DA 
as a particle-model solution to the Hubble tension. 
The axion couples to a dark non-Abelian gauge group\footnote{We have focused on SU(2). A generalization to SU(N) only changes numerical factors for $g_*$ and $\kappa$ without qualitative impact.}, 
which adds an additional friction to the equation of motion of the axion 
and sources a dark radiation bath as the field rolls down its potential. 
This overdamped system has a well understood UV-completion 
and greatly alleviates the fine-tuning concerns present for 
the scalar-field EDE solutions. 
The injection time 
and total amount of added energy content is 
quantified fully by two linear combinations of parameters:
$\frac{\Upsilon_0}{m^2}$ and $\frac{1}{2}m^2 \phi^2_0 \,$. 
The full theory has additional parameters, 
as the friction is determined by: 
$\Upsilon =\kappa \alpha^5 \frac{T^3_{\text{dr}}}{f^2}$. 
Here, $\kappa$ is an O(10) number, $\alpha < 0.1$,
$T_{\text{dr}} < f$, and $m \ll \alpha^2 T_{\text{dr}}$. 
For the sample values specified in the caption of
Fig.~\ref{fig:frac_energy_density}, we find that 
these conditions are easily satisfied for many different
combinations of viable parameters, for example: $m =4*10^{-25}\,$eV, $T_{\text{dr}}(z_{\text{peak}}) =0.4\,$eV, $f=0.3\,$GeV, 
$\alpha =0.1$, $\phi_0 = 10^{-3} M_{\text{Pl}}$, where $M_{\text{Pl}}$ is the reduced Planck scale.
We expect the full perturbative analysis to lift some of the degeneracy in these parameters and also in the choice of potential for the DA. 

We have solely investigated the overdamped \DA regime. Particle-sourcing friction could also play a role in an
underdamped regime.
Moreover, the \DA can be theorized to have a
UV-completion that ties its friction to the dark matter abundance. 
The symmetry breaking scale $f$ can, for example, be 
linked to the presence of heavy quarks charged under the dark SU(N). 
Thus, the dark matter abundance could be determined by $f$, 
which also controls the friction $\Upsilon$, 
potentially allowing a dynamical explanation for why the \DA 
begins to roll close to matter-radiation equality. 
We leave a detailed exploration of this to future work.
    
We note that $N_{\rm eff}$ constraints will not 
restrict this model. 
While the CMB was emitted at the redshift of recombination, 
the peaks of the CMB spectra in fact encode information 
from redshifts $z \lesssim 10^6$
\cite{Linder:2010wp, Karwal:2016vyq}. 
The \DA adds dark radiation to the Universe 
only after $\sim z_{\rm eq}$, unlike $N_{\rm eff}$ 
which adds radiation to the Universe at all times. 
Their imprints on the CMB peaks are hence different 
- the DA is expected to cause its largest change to the CMB 
close to the first peak in the TT spectrum based on Refs.~\cite{Linder:2010wp,Karwal:2016vyq}, while $N_{\rm eff}$ 
is not only constrained by matter-radiation equality, but also 
through its effect on the higher peaks in the CMB TT spectrum \cite{Follin:2015hya}. 
These distinct effects on the CMB imply that the DA model 
cannot be quantified by $N_{\rm eff}$, nor be restricted
by $N_{\rm eff}$ constraints. 

Lastly, we have invoked the \DA model
here as an explanation of extra dark energy components that 
resolve the Hubble tension, but this model 
has applications far beyond this tension. 
It has already been shown to be a viable candidate 
for cosmic inflation \cite{Berghaus:2019whh},
and could similarly drive the current cosmic acceleration (for example, \cite{Graham:2019bfu}). 
A family of scalar fields 
have often been theorized to cause the 
two known eras of cosmic expansion 
\cite{Kamionkowski:2014zda,Poulin:2018dzj}. 
We add the \DA to this list.

\begin{acknowledgments}
We are grateful to Marc Kamionkowski, David E. Kaplan, Vivian Poulin, Surjeet Rajendran, and Tristan Smith for discussions and feedback.
We acknowledge the support of NSF Grant No. PHY-1818899, NASA Grant NNX17AK38G and the Discovery Grant.
TK was also supported by funds provided by the Center for Particle Cosmology at the University of Pennsylvania. 
\end{acknowledgments}

 \bibliography{EDE}

\end{document}